\documentclass[a4paper]{llncs}
\usepackage[english]{babel}
\usepackage[utf8]{inputenc}
\usepackage{amssymb,amsmath}
\usepackage{url}
\usepackage{float}
\usepackage{graphicx}
\usepackage{multirow}
\usepackage{booktabs}

\begin{document}

\title{Effectiveness of pre- and inprocessing\\ for CDCL-based SAT
  solving} 
\author{Andreas Wotzlaw, Alexander van der Grinten, \and
  Ewald Speckenmeyer} 
\institute{Institut für Informatik, Universität zu  Köln,
  Pohligstr. 1, D-50969 Köln, Germany\\
\email{\{wotzlaw, vandergrinten, esp\}@informatik.uni-koeln.de}}

\maketitle

\begin{abstract}
  Applying pre- and inprocessing techniques to simplify CNF formulas
  both before and during search can considerably improve the performance
  of modern SAT solvers. These algorithms mostly aim at reducing the
  number of clauses, literals, and variables in the formula. However, to
  be worthwhile, it is necessary that their additional runtime does not
  exceed the runtime saved during the subsequent SAT solver
  execution. In this paper we investigate the efficiency and the
  practicability of selected simplification algorithms for CDCL-based
  SAT solving. We first analyze them by means of their expected impact
  on the CNF formula and SAT solving at all. While testing them on
  real-world and combinatorial SAT instances, we show which techniques
  and combinations of them yield a desirable speedup and which ones
  should be avoided.

  \keywords{satisfiability, preprocessing, inprocessing, CDCL solvers.}
\end{abstract}

\section{Introduction}
The satisfiability problem of propositional logic (SAT) has a number of
important real-world applications including hardware-verification,
software-verification, and combinatorial
problems~\cite{handbook}. Despite of being NP-complete,
real-world SAT instances can nowadays be solved in an acceptable time by
state-of-the-art solvers.

Among all approaches for SAT solving only {\em conflict-driven
  clause-learning} (CDCL) solvers~\cite{handbookCDCL}, an extension of
the DPLL procedure, have proved their remarkable efficiency in solving
real-world SAT problems, containing often more than 1 million variables
and more than 10 million clauses. It has been observed that their
performance can be improved if certain simplification techniques are run
on the CNF formulas before the actual SAT algorithm starts or during its
execution~\cite{bacchus03effectivepreprocessing,satelite,distillation,clause_elim_techs,unhiding,bce,jaervisalo2012inprocessing,vivification,niver}. Those
{\em preprocessing} and {\em inprocessing} techniques, respectively,
aiming mostly at reducing the number of clauses, literals, and variables
of the formula, have become an essential part of the SAT solving tool
chain. However, to be worthwhile, it is necessary that their additional
runtime does not exceed the runtime saved during the subsequent SAT
solver execution. It is a trade-off between the amount of reduction
achieved and invested time.
	
This paper evaluates some selection of promising pre- and inprocessing
techniques and combinations of them developed in the recent years.
Section~\ref{s:techniques} begins with preliminaries, describes briefly
simplification techniques considered here, and analyzes them by means of
their expected impact on the CNF formula and the CDCL-based SAT
solving. Section~\ref{s:evaluation} gives the evaluation of those
techniques and their combinations on real-world and combinatorial SAT
benchmarks. In Section~\ref{s:conclusion} we conclude our paper and give
recommendations, based on the measured effectiveness, for the usage of
simplification techniques for SAT solving.

\section{Pre- and Inprocessing Techniques}
\label{s:techniques}
For a Boolean variable $x\in V$, there are two literals, the positive
literal $x$ and the negative literal $\overline{x}$. A clause $C$ is a
disjunction of different literals over $V$ and a CNF formula a
conjunction of clauses. Let $\tau:V \rightarrow \{0,1\}$ be a truth
assignment. $\tau$ satisfies a literal $l$ iff $\tau(l)=1$. A clause
is satisfied by $\tau$ iff $\tau$ satisfies any of its
literals. $\tau$ satisfies formula $F$ iff all clauses of $F$ are
satisfied by $\tau$. A formula $F$ is {\em satisfiable} iff there is
at least one truth assignment satisfying it. Two formulas are {\em
  logically equivalent} if they are satisfied by exactly the same set
of assignments. A clause is a {\em tautology} if it contains both $x$
and $\overline{x}$ for some variable $x$.

\subsection{Selected Simplification Techniques} 
\label{ss:selection}
We give now a short description of the simplification techniques whose
evaluation we present in this paper. For a good overview of the recent
developments of pre- and inprocessing techniques for SAT solving we
refer to~\cite{biere2011preprocessing}.

{\em Subsumption} (SUB) is a simple technique trying to eliminate
logically redundant clauses from the CNF formula~\cite{handbook}. In its
simplest form SUB removes a clause $C$ iff there is a clause $D$ that
contains all literals in $C$ as a subset. For instance, for the formula
$F = (\neg x_1 \lor x_2) \land (\neg x_1 \lor x_2 \lor x_3) \land (\neg
x_1 \lor \neg x_2)$, SUB will remove the second clause since the first
clause is its proper subset. SatELite~\cite{satelite} introduces another
subsumption variant. Here by resolving clauses $(\neg x_1 \lor x_2)$ and
$(\neg x_1 \lor \neg x_2)$ in $F$, we obtain the new clause $(\neg x_1)$
subsuming both former clauses. The algorithm trying to resolve two
clauses to find a resolvent subsuming both input clauses is called {\em
  resolution subsumption} (RSUB), or {\em self-subsuming
  resolution}~\cite{satelite}, and constitutes another important
simplification technique.

{\em Bounded variable elimination} (BVE)~\cite{satelite,niver} uses
the Davis-Putnam procedure~\cite{handbook} to remove variables from
the formula. It chooses first a variable and then removes all clauses
containing this variable from the formula while adding to the formula
all resolvents of those clauses with respect to the variable
chosen. To prevent exponential blow-up of the formula, variables are
only eliminated if the number of new clauses is less than the number
of removed clauses.

We call a clause $C$ {\em blocked} if there is a literal $l$ in $C$ so
that all resolvents between $C$ and any other clause with respect to the
variable inducing $l$ are tautologies. Such a blocked clause can be
removed from the formula without affecting the satisfiability properties
of the formula. The procedure that removes all blocked clauses from the
formula is called {\em blocked clause elimination} (BCE) and was
introduced as a preprocessing technique in~\cite{bce}.

{\em Unhiding} (UH)~\cite{unhiding} is a technique performing a depth
first search on the binary implication graph~\cite{bin_reasoning} formed
by the clauses of length two of the CNF formula. It finds all strongly
connected components (SCC) of the graph as well as a subset of the
failed literals~\cite{handbook}, and transitive edges that can be
removed without affecting the satisfiability of the formula. Here all
literals represented by the nodes of an SCC are equivalent and can be
replaced by a single one. During the search various information is
extracted that can be used both for {\em hidden literal elimination}
(HLE)~\cite{unhiding} and {\em hidden tautology elimination}
(HTE)~\cite{clause_elim_techs}.

{\em Distillation} (DI) is like failed literal
elimination~\cite{berre01exploiting} a probing based technique. Consider
the clause $(l_1 \lor \ldots \lor l_k)$ for a fixed order of its
literals. Next assign sequentially all literals one after the other to
$0$ according to the order chosen and perform {\em unit propagation}
(UP) after setting each literal. This will either lead to a conflict, or
there will be a literal $l_i$ that has already been set by UP before it
is set to $0$ by the procedure. In both cases we can try to shorten the
clause by removing one or more of its literals. There are multiple
variants of distillation and we refer
to~\cite{distillation,vivification} for more details.

\subsection{Properties of Simplification Techniques}
We state now some criteria according to which we assess the impact of
the algorithms described above on the CNF formula and the SAT solving.
\begin{enumerate}
\item\textbf{Preservation of unit propagation.} CDCL solvers rely on UP
  to fix variables while expanding the search tree. It is desirable that
  simplification techniques preserve UP, i.e., if a variable $v$ can be
  fixed to a certain value when applying UP to the input formula and a
  given partial truth assignment $\tau$, then the same variable should
  be fixed by UP when applied to the simplified formula and the same
  assignment $\tau$.
\item\textbf{Preservation of equivalence.} If an algorithm does not
  preserve the logical equivalence, it is often essential to construct a
  satisfying assignment of the input formula from a satisfying
  assignment of the simplified formula.
\item\textbf{Simulation by resolution.} Some applications of SAT solvers
  rely on resolution refutation proofs for unsatisfiable instances,
  e.g., partial MAX-SAT solvers based on iterative SAT
  solving~\cite{fu2006onsolving}. Hence it should be possible to
  construct efficiently a resolution refutation of the input formula
  given a resolution refutation for the formula resulting from the
  simplification.
\item\textbf{Confluence.} We call an algorithm {\em confluent} if its
  result does not depend on the order in which variables and clauses of
  the input formula are inspected. A non-confluent algorithm may need
  additional heuristics to improve the order in which variables or
  clauses are processed.
\item\textbf{Implementation.} It is desirable that the preprocessing
  algorithm can be implemented with data structures that are already
  present in CDCL solvers. Algorithms that do not rely on special data
  structures may also be better suitable for inprocessing in addition
  to preprocessing. Some techniques require data structures which for
  each literal list all clauses containing it. Those so called {\em
    literal occurrence lists} are usually not required for the CDCL
  algorithm and maintaining them would result in a performance
  penalty.
\end{enumerate}
\begin{table}
  \centering
  \caption{Properties of the simplification techniques (*: modulo
    variable renaming).}
  \begin{tabular}{lccc}\toprule
    Preprocessing & Preserves & Preserves & Requires \\
    Technique & UPs & Equivalence & Occ-Lists \\ \midrule
    (Resolution-) Subsumption ((R)SUB) & Yes & Yes & Yes \\
    Bounded Variable Elimination (BVE) & -- & -- & Yes \\
    Blocked Clause Elimination (BCE) & -- & -- & Yes \\
    Hidden Tautology Elimination (HTE) & -- & Yes & -- \\ 
    Hidden Literal Elimination (HLE) & Yes & Yes & -- \\
    Unhiding without HLE/HTE (UH) & Yes & Yes* & -- \\
    Distillation (DI) & Yes & Yes & -- \\\bottomrule
  \end{tabular}
  \label{table:simp_props}
\end{table}

Table~\ref{table:simp_props} lists properties of the simplification
algorithms described in Section~\ref{ss:selection} according to the
criteria given above. Additionally, all algorithms can be simulated by
resolution whereas none of them, except for BCE, is confluent.

\section{Comparative Evaluation}
\label{s:evaluation}
\subsection{Experimental Setup}
The simplification techniques presented in Section~\ref{s:techniques}
were implemented in our sequential CDCL-based solver {\em
  satUZK}~\cite{satuzk} having performance comparable with that of
MiniSAT 2.2~\cite{minisat}. All tests were run on a machine with two
Intel Xeon E5410 2.33 GHz processors running a 64-bit Linux 2.6.32
with 32GB RAM.

There are two categories of instances that were tested: {\em
  application} and {\em hard combinatorial} instances. Our test suits
were formed from a subset of all instances from the SAT Challenge
2012~\cite{challenge_2012}. First, instances which could be solved with
MiniSAT 2.2 on our test machine in between 120 and 600 seconds were
selected, resulting in 58 hard combinatorial instances and a large
number of application instances, from which we took 60, selected
uniformly at random.  For each of the 118 CNF formulas five instances
were generated at random by permuting the orders of clauses, the orders
of variables inside each clause, and the polarities of the variables,
resulting in 300 application and 290 hard combinatorial test
instances. This was done in order to test the robustness of the
simplification techniques as well as to improve the reliability of the
results.

The timeout for solving each instance was set to 600 seconds. The solver
was allowed to use 10\% of the timeout on preprocessing and 10\% of the
timeout on inprocessing. These values have been determined empirically
through testing. Inprocessing was run each time the ratio of the
inprocessing time-limit consumed so far and the current solver runtime
was less than $0.1$. In addition to testing each simplification
technique separately, we tested combinations of them, e.g., the
combination of BCE and BVE, denoted in the following by BCE+BVE. This
notion specifies also the order in which the techniques of the
combination were applied. To prevent the solver from wasting time on
instances where preprocessing had little effect, we used the following,
empirically determined, heuristic:
\begin{enumerate}
\item Perform a single round of preprocessing.
\item If more than 1\% of the remaining variables could be eliminated in
  1\% of the timeout available for solving the instance, then go to
  1. and start another round. Otherwise, stop preprocessing.
\end{enumerate}

Table~\ref{table:res_app} reports on the effects of the specified
simplification techniques, called here {\em configurations}, in the
application and hard combinatorial categories. Here only configurations
giving the best results are presented whereas for inprocessing only
lightweight techniques like UH and DI not requiring literal occurrence
lists were tested. The reference configuration invokes the CDCL
algorithm implemented in satUZK without doing any pre- or inprocessing
at all. Note that in contrast to the original instances, it was now not
possible to solve all permuted instances within the timeout. Moreover,
the RSUB configuration also performs SUB in addition to RSUB, whereas UH
does unhiding combined with HLE and HTE based on the information
extracted during the unhiding phase. Finally, DI is applied only to the
100 most active clauses during inprocessing. Here, the variables of a
clause are assigned in order of decreasing VSIDS~\cite{handbookCDCL}
activity.
\begin{table}
  \centering
  \caption{Statistics on the application and combinatorial results.
    The average percentage of variables ($\Delta$vars) and clauses
    ($\Delta$cls) eliminated per instance by pre- and inprocessing are
    given in columns 4 and 5, and 7 and 8, respectively. Note that BVE
    was allowed to add additional but valuable binary clauses. (*: RSUB
    and BCE do not remove or fix variables, **: not computed for
    inprocessing).}
  \begin{tabular}{llcccccc}
    \toprule
    && \multicolumn{3}{c}{Application} &
    \multicolumn{3}{c}{Combinatorial}\\
    \cmidrule{3-5}\cmidrule{6-8} 
    Type&Configuration& solved & $\Delta$vars[\%] & $\Delta$cls[\%]
    & solved &    $\Delta$vars[\%] & $\Delta$cls[\%]\\
    \midrule
    Reference & -- & $184$ & -- & -- & $185$ & -- & -- \\
    \cmidrule{2-8}
    \multirow{7}{*}{Preprocessing} 
    & BVE & $224$ & $-39.59$ & $+19.11$ & $178$ &$-16.67$ & $+0.53$\\
    & RSUB & $164$ & * &  $-0.18$ & $182$ & * & $-1.75$ \\
    & BCE & $192$ & * &  $-0.31$ & \textbf{190} & * & $-0.78$ \\
    & UH & $200$ & $-10.36$ & $-2.80$ & $187$ & $-4.27$ &$-2.70$ \\
    & BCE+UH & $203$ & $-10.37$ & $-3.37$ & $189$&$-4.27$ & $-3.45$\\
    & BCE+BVE & $217$ & $-40.08$ & $+17.99$ & $184$ & $-18.14$&$-0.36$\\
    & BCE+BVE+UH&\textbf{233}&$-49.63$ & $+24.78$& $182$ &$-19.12$ &$-1.40$\\
    \cmidrule{2-8}
    \multirow{3}{*}{Inprocessing} 
    & UH & $191$ & ** & ** & $183$ &**&**\\
    & DI & $188$ & ** & ** & $178$ & ** & ** \\
    & UH+DI & \textbf{202} & ** & ** & $183$ & ** & ** \\
    \bottomrule
  \end{tabular}
  \label{table:res_app}
\end{table}

\subsection{Results for Application Instances}
\begin{figure}
  \centering
  \includegraphics[scale=.803]{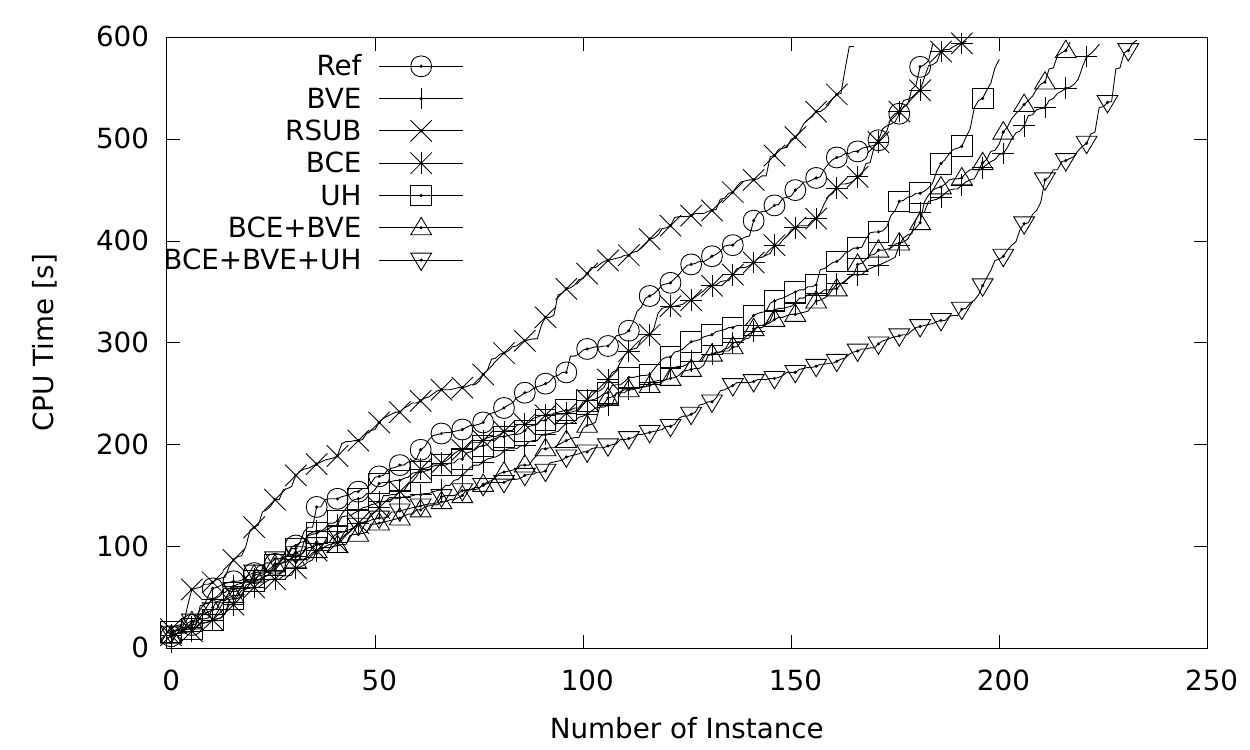}
  \caption{Effects of preprocessing for solving application instances.}
  \label{fig:app_pre_cactus}
\end{figure}

Figure~\ref{fig:app_pre_cactus} shows a cactus plot of different
preprocessing configurations used for the permuted application
instances. The x- and y-axis report on the instance number and the CPU
time (in seconds) required to solve that instance, respectively.
Instances are ordered by increasing runtime and counted with 600 seconds
if not solved. All preprocessing configurations but RSUB perform better
than the reference configuration. Configuration BVE leads to the highest
performance improvement executing a single technique only. Applying BVE
improved both the runtime and the number of solved instances. UH
produces good results as well. The results could be further improved by
combining BCE, BVE, and UH. Here, BCE alone seems to have a slightly bad
effect on the solver's performance probably because BCE does not
preserve UP.
\begin{figure}
  \centering
  \includegraphics[scale=.803]{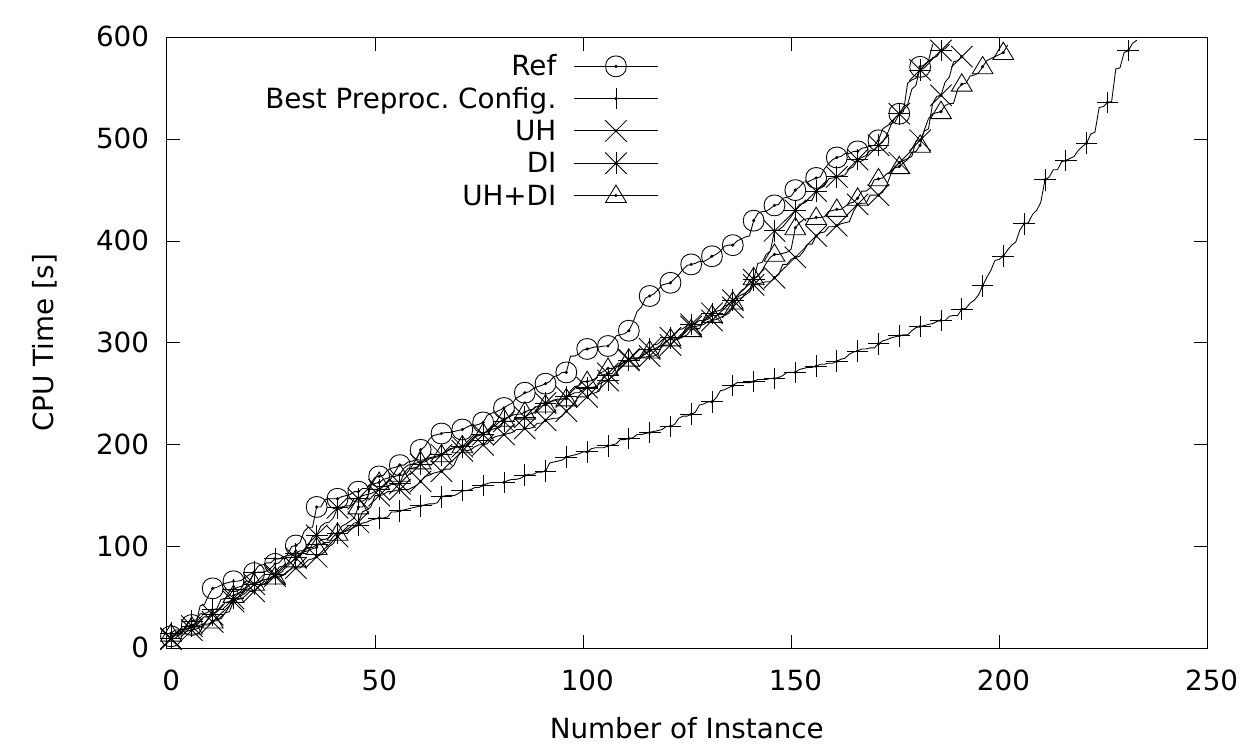}
  \caption{Effects of inprocessing for solving application instances.}
  \label{fig:app_in_cactus}
\end{figure}

Furthermore, the bad performance of RSUB for application instances, may
be explained by many cache-misses caused by subsumption when iterating
through literal occurrence lists of the CDCL Solver. Interestingly,
subsumption performs much better when applied to instances that have not
been permuted, possibly because the variables are not randomly
distributed across those instances. To assure that the results obtained
did not originate from our subsumption implementation, we rerun the
tests using MiniSAT 2.2 with similar results.

According to Figure~\ref{fig:app_in_cactus} the performance improvements
achieved by inprocessing alone for solving application instances are
smaller than those obtained with preprocessing. Here UH appears to be
the best configuration executing a single technique only, followed by DI
solving only four more instances than the reference configuration. The
best effects on SAT solving shows here UH+DI.

\subsection{Results for Hard Combinatorial Instances}
\begin{figure}
  \centering
  \includegraphics[scale=.8]{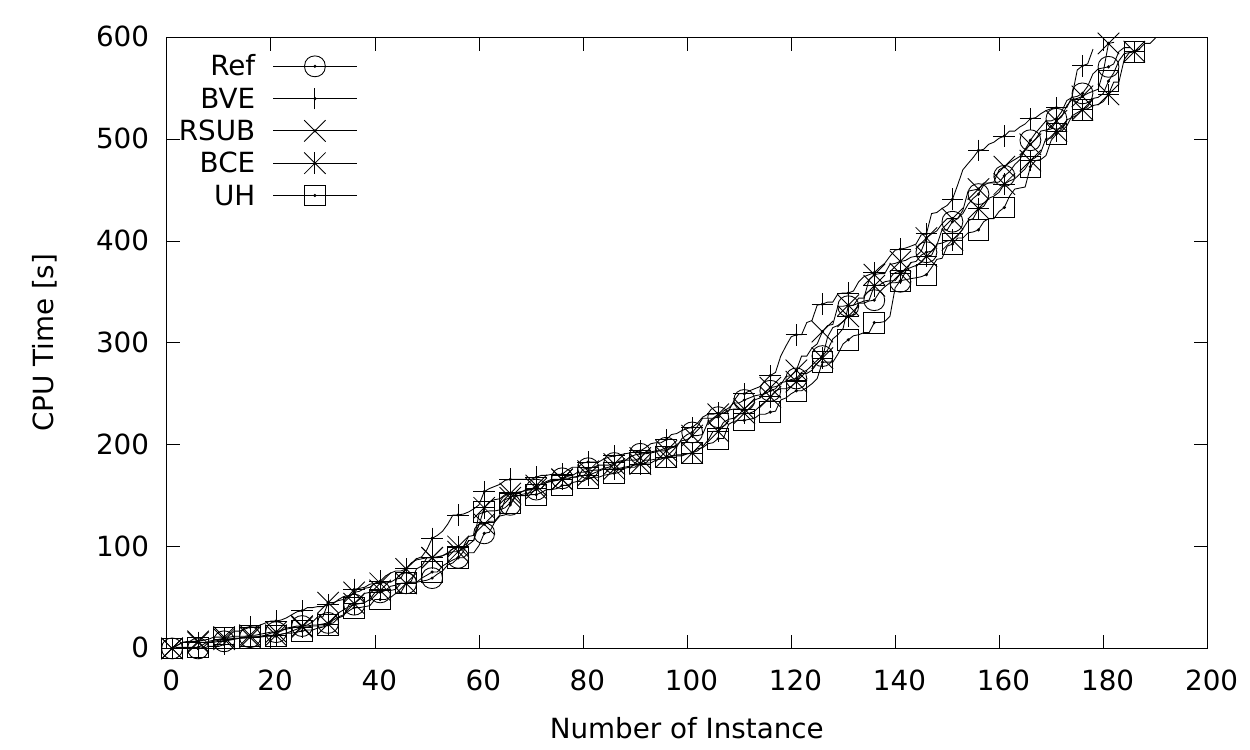}
  \caption{Effects of preprocessing for solving hard combinatorial
    instances.}
  \label{fig:crafted_pre_cactus}
\end{figure}

\begin{figure}
  \centering
  \includegraphics[scale=.8]{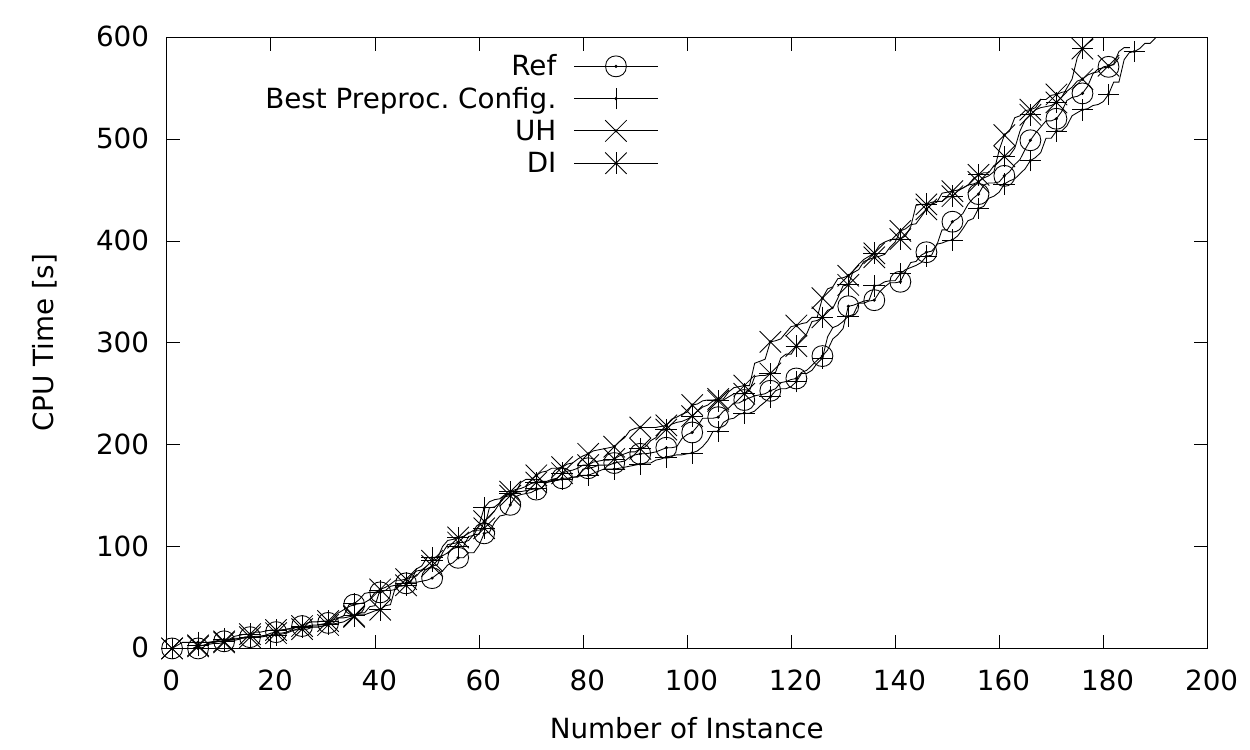}
  \caption{Effects of inprocessing for solving hard combinatorial
    instances.}
  \label{fig:crafted_in_cactus}
\end{figure}

Figure \ref{fig:crafted_pre_cactus} shows the effects of the different
preprocessing techniques on the permuted combinatorial instances. Here
only BCE and UH, both single-technique categories, outperform the
reference configuration. We suspect that the reason for that is that UH
improves UP by removing hidden literals. Moreover, opposite to the
application category, the combination of multiple simplification
algorithms does not yield better results than BCE alone.  Other
techniques like BVE or BCE+BVE can even prevent the solver from doing
useful UP by removing clauses required for propagations. As hard
combinatorial instances are small compared to application instances, the
probability that important clauses are deleted is higher. Additionally,
the results from Table~\ref{table:res_app} show that both BVE and UH
remove fewer variables on hard combinatorial instances than they do on
application instances. This, together with
the inprocessing results shown in Figure~\ref{fig:crafted_in_cactus},
indicates that simplification techniques are not as beneficial for
solving hard combinatorial instances as they are for real-world
application instances.

\section{Conclusion}
\label{s:conclusion}
In this paper the effectiveness of selected simplification techniques
and combinations of them for modern CDCL-based SAT solving has been
examined. The techniques selected were first assessed on the basis of
their theoretical characteristics important for SAT solving like
preservation of UP and logical equivalence, simulatability of
resolution, and implementation-related issues like practicable running
times, or requirement of additional expensive data structures.

Applying preprocessing techniques to simplify CNF formulas coding
real-world problems has proved to be extremely beneficial for the
performance of the SAT solver. Here the combination of BCE, BVE, and
UH was the most effective preprocessing configuration, followed by BVE
and BCE+BVE. With proper preprocessing techniques it was possible to
solve up to $27$\% more instances.

The inprocessing was generally not as effective as preprocessing. For
solving application instances, UH+DI was the most effective
configuration, improving acceptably over the reference
solution. However, for the combinatorial category both UH and DI were
not so successful. This indicates that developing effective
inprocessing techniques is non-trivial and their success depends much
on the instance type. It requires in-depth knowledge about how
different techniques can be combined and integrated efficiently into
the solvers' search procedure.

Finally, simplification techniques have much more effect for solving
real-world SAT problems than for hard combinatorial instances, where
their application could even be counterproductive. 

\bibliographystyle{splncs03}
\bibliography{literature}

\end{document}